Centre de Physique Théorique[*] - CNRS - Luminy, Case 907
F-13288 Marseille Cedex 9 - France

# Non-Gaussian statistics in space plasma turbulence : fractal properties and pitfalls

T. Dudok de Wit[1] and V.V. Krasnosel'skikh[2]

Abstract

Statistical properties of collisionless plasmas in the vicinity of the Earth's bow shock have been investigated with the aim to characterize the intermittent behaviour of non-magnetohydrodynamic turbulence. The structure functions of the fluctuating magnetic field reveal an increasing departure from Gaussianity at small scales, which is similar to that observed in solar wind turbulence and is surprisingly little affected by the abrupt shock transition. While these results may be the signature of a multifractal process, a deeper inspection reveals caveats in such an interpretation. Several effects, including the anisotropy of the wavefield, the violation of the Taylor hypothesis and the occasional occurrence of coherent wave packets, strongly affect the higher order statistical properties. Most of the small differences observed between the up- and downstream sides of the shock can be ascribed to the occurrence of discrete whistler wavetrains, while the wavefield itself is much less intermittent. It is also shown how the finite length of the records prohibits a reliable estimation of structure functions beyond the fourth order. These results preclude an unambiguous identification of underlying models for intermittency.

---

[*]  Unité Propre de Recherche 7061
[1]  also at the Université de Provence;  E-mail : ddwit@cpt.univ-mrs.fr
[2]  Laboratoire de Physique et Chimie de l'Environnement, CNRS, 3A Av. de la Recherche Scientifique, 45071 Orléans cedex 2, France

# Non-Gaussian statistics in space plasma turbulence: fractal properties and pitfalls

T. Dudok de Wit[1] and V.V. Krasnosel'skikh[2]

[1] Centre de Physique Théorique, CNRS and Université de Provence, Luminy case 907, 13288 Marseille cédex 9, France
[2] Laboratoire de Physique et Chimie de l'Environnement, CNRS, 3A, Av. de la Recherche Scientifique, 45071 Orléans cédex 2, France

**Abstract**. Statistical properties of collisionless plasmas in the vicinity of the Earth's bow shock have been investigated with the aim to characterize the intermittent behaviour of non-magnetohydrodynamic turbulence. The structure functions of the fluctuating magnetic field reveal an increasing departure from Gaussianity at small scales, which is similar to that observed in solar wind turbulence and is surprisingly little affected by the abrupt shock transition. While these results may be the signature of a multifractal process, a deeper inspection reveals caveats in such an interpretation. Several effects, including the anisotropy of the wavefield, the violation of the Taylor hypothesis and the occasional occurrence of coherent wave packets, strongly affect the higher order statistical properties. Most of the small differences observed between the up- and downstream sides of the shock can be ascribed to the occurrence of discrete whistler wavetrains, while the wavefield itself is much less intermittent. It is also shown how the finite length of the records prohibits a reliable estimation of structure functions beyond the fourth order. These results preclude an unambiguous identification of underlying models for intermittency.

## 1 Introduction

In the last decades, large amounts of fluctuation data have been gathered in the interplanetary medium, partly with the aim to test theoretical predictions on plasma turbulence (Grappin et al., 1991). These data have greatly improved our understanding and revealed some striking similarities with neutral fluids. The spectral index, for example, is often found to be closer to the celebrated -5/3 constant predicted by Kolmogorov for the inertial range of hydrodynamic turbulence (e.g. Frisch, 1995) than to the -3/2 prediction from magnetohydrodynamic (MHD) theory (Kraichnan, 1965). More generally, space plasmas exhibit the same intermittent behaviour and deviation from Gaussian statistics at small scales as observed in neutral fluid turbulence. This means that second order moments such as Fourier power spectra do not completely describe the wavefield and hence higher order moments must be investigated.

Among the possible approaches to the characterization of non Gaussian processes, the structure function has received wide attention since it can conveniently be cast into a theoretical framework (e.g. Monin and Yaglom, 1975). Numerous successful applications to fluid turbulence have already been reported (e.g. Frisch, 1995). Its application to plasma turbulence has been pioneered by the work of Burlaga et al. (1986; 1991a; 1991b; 1992; 1993) and later followed by other studies (Marsch and Liu, 1993; Marsch and Tu, 1994; Carbone, 1993; 1994; Carbone et al., 1995; 1996; Ruzmaikin et al., 1995), all of them based on solar wind data. These studies have shown evidence for intermittency, with an increasing departure from Gaussian statistics at small scales. The degree of intermittency, however, often varies within the same medium. In solar wind turbulence, for example, a marked transition appears between slow and fast wind conditions (Marsch and Liu, 1993).

The scaling of higher order moments in turbulent wavefields has generally been interpreted in terms of fractal or multifractal processes (Paladin and Vulpiani, 1987) but their physical meaning for plasma turbulence remains unclear. Indeed, magnetofluids are often highly structured and therefore much less amenable to a comprehensive theoretical description as neutral fluids with the Navier-Stokes equation. An important difference results from the large-scale magnetic field that is responsible for the Alfvén effect (Biskamp, 1993). The magnetic field also contributes to make the turbulent wavefield anisotropic and allows different types of waves to coexist. This relative complexity of plasma turbulence constitutes a major obstacle to a more quantitative understanding of its statistical properties. It does not prevent, however, such statistical analyses from providing new insight on occasion. By comparing different regions of the solar wind, for example, Marsch and Tu (1994) were able to show that some of them were not just remnants of coronal events but rather consisted of dynamically evolving MHD turbulence.

In view of these results, we are motivated to compare statistical properties of space plasma turbulence in the vicinity of the Earth's collisionless bow shock. This is a region of special interest since it offers a paradigm for nonlinear effects in strong plasma turbulence; it also provides a fine example of a sudden transition (Stone and Tsurutani, 1985). In particular, the turbulence in that region is an inherent and important element of the particle acceleration process. In contrast to solar wind turbulence, it is strongly dispersive and cannot be considered as magnetohydrodynamic (MHD) since the time scales under consideration are of the same order as the inverse gyrofrequencies of the different ion species. Nevertheless, most of the statistical concepts developed for the inertial range of hydrodynamic or MHD turbulence remain applicable, using symmetry properties of the nonlinear interactions (Zakharov, 1984). An additional feature of the turbulent wavefield is its strong interaction with the energetic particles that are accelerated at the bow shock. Since the dominant physical processes involved on the two sides of the shock front are likely to be quite different, we are particularly interested in comparing their statistical properties with the aim to assess possible invariants. Incidentally, such an investigation reveals several pitfalls that often seem to be overlooked in the interpretation of higher order statistical moments. It is shown how different effects such as the anisotropy of the wavefield, the occurrence of discrete wave packets and the finite duration of the records preclude a straightforward interpretation of the higher order moments.

turbulence, however, is the presence of energetic ions that escape from the shock front and propagate upstream. Furthermore, as mentioned before, the turbulence is not of the MHD type. The most energetic ions interact with the wavefield, causing certain waves to steepen and evolve into large-amplitude monolithic structures called shocklets (Thomson et al., 1990) or Short Large-Amplitude Magnetic Structures (SLAMS) (Schwartz et al., 1992). These nonlinear structures in turn scatter the ions and have received wide interest because of the role they play in the dynamics of the shock formation. The downstream side of the shock is in contrast characterized by weaker nonlinear interactions as the energetic tail of the ion distribution merely contributes to a heating of the bulk. It is also closer to equilibrium whereas the upstream turbulence is driven by wave-particle interactions. Thus, we have an abrupt transition between two different regions whose macroscopic properties strongly differ. Such a transition offers a unique opportunity for comparing statistical properties of the wavefield.

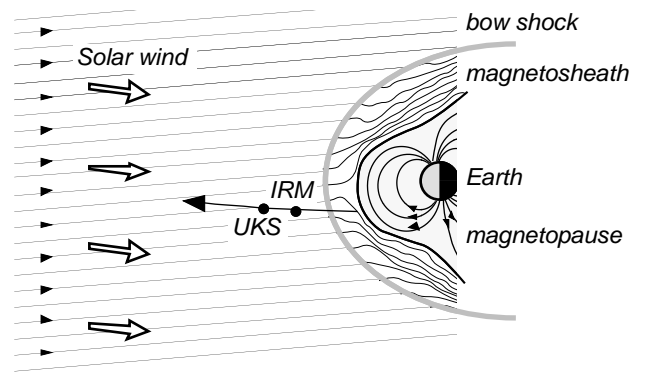

**Fig. 1.** Schematic representation of the Earth's bow shock with the magnetic field lines, the orientation of the solar wind and the configuration of the satellites for the event considered in this paper. The sun is situated on the left.

## 2 Experimental context

Our investigation is based on magnetic field data that have been gathered by the AMPTE-UKS satellite at the Earth's quasi-parallel bow shock on day 304 of 1984. This event has been described more in detail by Schwartz et al. (1992) and by Mann et al. (1994). Figure 1 illustrates the configuration of the shock front which is located at the sunward side of the Earth. The standing shock wave results from a sudden deceleration of the supersonic solar wind at the encounter of the Earth's magnetosphere. The resulting compression modifies the plasma parameters and causes a sudden increase in the pressure. Both sides of the bow shock are qualified as strongly turbulent because the relative fluctuation levels are large.

The turbulence in the upstream (i.e. sunward) region essentially consists of a mixture of waves and structures that are convected towards the Earth by the solar wind. A review on the latter has recently been given by Tu and Marsch (1995). A basic difference with respect to solar wind

The magnetic field data of interest consist of two continuous records of 9600 and 8867 samples gathered respectively in the up- and downstream regions with a sampling rate of 8 Hz. For each record we have three components of the magnetic field labeled $B_x$, $B_y$ and $B_z$. Upstream the bow shock, the solar wind velocity and the prevailing magnetic field are almost parallel and make an angle of about 15° with $B_x$. We checked that spurious effects such as aliasing, lack of dynamic range and a nonlinear response of the fluxgate magnetometers (Southwood et al., 1985) are not important for what follows.

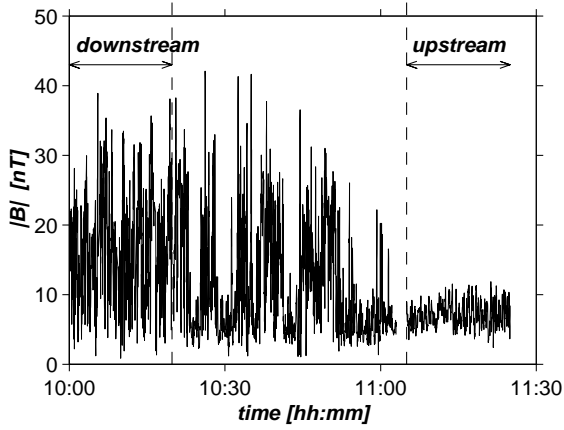

**Fig. 2.** Time evolution of the magnetic field amplitude measured by the AMPTE-UKS spacecraft as it crosses the bow shock on an outbound portion of its orbit. Three sequences can be distinguished: the stationary field up- and downstream the bow shock, and a transition region. The sequences considered in this paper are marked on the figure.

Figure 2 displays the time evolution of the wavefield as the AMPTE-UKS spacecraft crosses the bow shock on a outbound orbit. The most evident manifestation of this crossing is a sudden decrease of the magnetic field fluctuation level by a factor of about five. The shock transition is non monotonic and non stationary and therefore gives the impression to be crossed several times before the spacecraft definitely leaves the magnetosheath. Our two records have been selected sufficiently far away from the shock transition to guarantee stationarity. One of the basic properties of the wavefield, expressed in Fig. 3, is its total power spectral density. We should point out that these results and the ones that follow are all expressed in the spacecraft frame of reference, in which the frequencies are Doppler-shifted by to the solar wind flow.

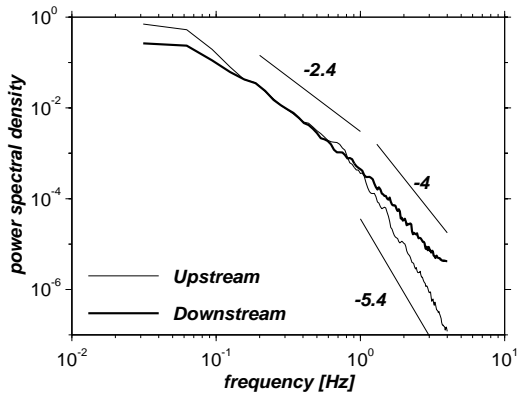

**Fig. 3.** Total power spectral density of the magnetic field calculated for the up- and downstream regions. The amplitudes have been adjusted to ease comparison and some spectral indices are indicated. Note that the frequencies are measured in the spacecraft frame of reference.

The power spectra in Fig. 3 reveal that both sides of the shock are characterized by two spectral indices, each of which spans approximately one decade. The presence of two spectral indices is common in interplanetary plasmas and strictly speaking excludes scale invariance. Our objective now is to consider higher order moments of the data at different scales, and see if these are related to macroscopic properties.

## 3 Higher order moments

The classical tool for quantifying higher order statistical properties of a velocity field $v(x)$ at a scale $l$ is the q'th order structure function (Monin and Yaglom, 1975)

$$S_q(l) = \left\langle |v(x+l) - v(x)|^q \right\rangle , \quad (1)$$

in which brackets denote ensemble-averaging. In the inertial range of turbulence, the energy transfer rate between nonlinearly interacting eddies is constant and a universal scaling law holds

$$S_q(l) \propto l^{\alpha(q)} , \quad (2)$$

where for Gaussian statistics and a spectral density that scales as $P(k) \propto k^{-\gamma}$ the scaling exponent is related to the spectral index $\gamma$ by

$$\alpha(q) = \frac{q}{2}(\gamma - 1) . \quad (3)$$

Conversely, any deviation from this linear relation suggests that the turbulence is intermittent and thus departs from a Gaussian process. For practical purposes, we also introduce the dimensionless structure function

$$A_q(l) = \frac{S_q(l)}{S_2(l)^{q/2}} \propto l^{\beta(q)} . \quad (4)$$

For Gaussian scale-invariant turbulence, its scaling exponent $\beta(q) = \alpha(q) - \alpha(2)q/2$ is equal to zero and hence $A_q(l)$ does not depend on $l$. Negative values of the scaling exponent in turn imply that the tails of the probability distribution are enhanced with respect to a normal distribution. Following Rose and Sulem (1978), we can relate $\beta(q)$ to the Kolmogorov capacity or fractal dimension D of the turbulent eddies

$$\beta(q) = (3-D)\frac{2-q}{2} . \quad (5)$$

This fractal dimension $D \leq 3$ expresses how space-filling the turbulent eddies are; D=2 for example corresponds to vortices that are stretched into sheet-like structures.

These concepts are usually formulated in the context of non-dispersive media but remain essentially valid in dispersive media when the energy spectrum is determined by nonlinear processes. Decay instabilities, for example, lead to dispersion relations that satisfy the scale invariance $\omega(\varepsilon k) = \varepsilon^\alpha \omega(k)$, where $\omega$ is the angular frequency and k the wavenumber. A similar scale invariance holds for the matrix elements $V(\varepsilon k, \varepsilon k_1, \varepsilon k_2) = \varepsilon^\beta V(k, k_1, k_2)$ that describe the nonlinear wave interactions (Zakharov, 1984). If the energy input of the turbulent wavefield occurs at large scales and the damping at small scales, with a sufficiently large interval between the two, then the physical processes in the inertial range are governed by nonlinear wave interactions. Assuming that the spectral energy flux is constant, one then recovers a power spectral density of the form $P(k) \propto k^{-\gamma}$, where the spectral index $\gamma$ is determined by the self-similarity exponents $\alpha$ and $\beta$. Clearly, these properties only hold if the scales under consideration are located sufficiently far away from the generation and dissipation processes. In our case and for the upstream region, most of the energy input occurs in a well defined frequency band centered on $\omega/2\pi = 0.1$ Hz (Dudok de Wit and Krasnosel'skikh, 1995). For that reason, we'll hereafter focus on time scales that are smaller than 10 sec.

The first problem that arises in applying the structure function approach to experimental data is its spatial dependence, which most experiments cannot properly resolve. This problem is generally overcome by invoking the Taylor hypothesis, which states that the intrinsic time dependence of the wavefield can be ignored when the turbulence is convected past the probes at nearly constant speed. With this hypothesis, the temporal dynamics should reflect the spatial one. Furthermore, since we measure magnetic field rather than velocity fluctuations, we set

$$S_q(\tau) = \left\langle |B(t+\tau) - B(t)|^q \right\rangle , \qquad (6)$$

where brackets now denote time-averaging, assuming ergodicity. For MHD turbulence, this new expression is a good approximation of Eq. (1) since the wavefield satisfies

$$\frac{d\mathbf{v}}{dt} = -\frac{1}{4\pi\rho}(\mathbf{B}\cdot\nabla)\mathbf{B} , \qquad (7)$$

where $\rho$ is the mass density of the magnetofluid. If the turbulence is frozen in the convective flow then

$$|\mathbf{B}| \propto \rho . \qquad (8)$$

Inserting this into Eq. (7) and taking the Fourier transforms in wavenumber **k** and angular frequency $\omega$ then gives

$$\omega|\mathbf{v}| \propto \mathbf{k}\cdot\mathbf{B} . \qquad (9)$$

This expression attests the existence of a one-to-one relation between the spatial structure of $\mathbf{v}(\mathbf{x})$ and the temporal dynamics of $\mathbf{B}(t)$, provided that the wavefield is dispersionless, namely for $\omega \propto |\mathbf{k}|$. Many experimental studies on space plasma turbulence tacitly make this assumption which considerably eases the analysis of time series. Yet, little work has been done so far to check the validity of the Taylor hypothesis in space plasmas. A reason for this is the lack of spatio-temporal data, for which multiple-spacecraft missions are necessary. The AMPTE mission is a notable exception since the AMPTE-UKS satellite was closely followed by a second spacecraft, AMPTE-IRM. This offered a unique opportunity to estimate the dispersion relation (Dudok de Wit et al., 1995). Figure 4 reveals the measured dispersion relation, showing that it is essentially linear downstream but nonlinear upstream.

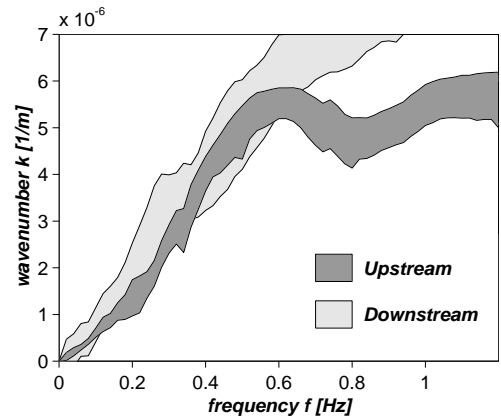

**Fig. 4.** The dispersion relation of the magnetic field as measured up- and downstream the bow shock; k stands for the projection of the wavenumber along the spacecraft separation vector. The width of the bands reflects the scatter in the frequency-wavenumber spectrum.

Clearly, such a nonlinear dispersion invalidates the Taylor hypothesis. More importantly, it implies that the statistical properties of $\mathbf{B}(t)$ and $\mathbf{v}(\mathbf{x})$ are too loosely connected to establish a quantitative link between them. Their spectral indices, for example, may differ. We'll therefore restrict our attention to time scales of the magnetic field only, deliberately avoiding any extrapolation to the properties of the velocity field itself.

Before investigating the structure function, it is informative to visualize the probability distribution of the fluctuation amplitudes. Figure 5 displays the probability density of the differences $\delta B(\tau) = B(t+\tau) - B(t)$ for different values of $\tau$. A lack of self-similarity is already apparent, as the density continuously evolves from a Gaussian distribution (at large $\tau$) to a mixture between a Gaussian and a Laplacian distribution (at small $\tau$).

The dimensional and dimensionless structure functions of the magnetic field component $B_x$ are displayed in Fig. 6. The structure functions of the other components are qualitatively similar and therefore not represented. Figures 5 and 6 confirm what has been observed before in solar wind plasmas (Burlaga, 1991b; Marsch and Liu, 1993; Marsch and Tu, 1994), namely that the turbulence becomes increasingly intermittent at small time or spatial scales. Intermittency manifests itself here as the presence of large-amplitude bursts that occur erratically in time. This property is best revealed by the dimensionless structure function, which is not constant. We also show for comparison results obtained from a surrogate version of the upstream data. The surrogate data mimic the linear properties of the original data, such as the power spectral density, but are by construction Gaussian at all scales (Theiler et al., 1992). One can thereby easily check that for surrogate data $\beta(q) \equiv 0$ and hence the intermittency observed in the original data is significant.

A surprising result here is the apparent similarity between the non Gaussian features as they are observed up- and downstream the bow shock. One would indeed expect a more marked difference given their differing macroscopic properties. This similarity may indicate that the small scale structure of the turbulence is not so much affected by the abrupt shock transition. We must stress, however, that the structure function is not necessarily the most appropriate tool for distinguishing the two regimes. In particular, more contrasted pictures have been obtained by using other quantities such as higher order Fourier spectra (Dudok de Wit and Krasnosel'skikh, 1995). Thus, a qualitative inspection of the structure functions in Fig. 6 isn't informative enough as far as differences are concerned and a finer analysis must be carried out.

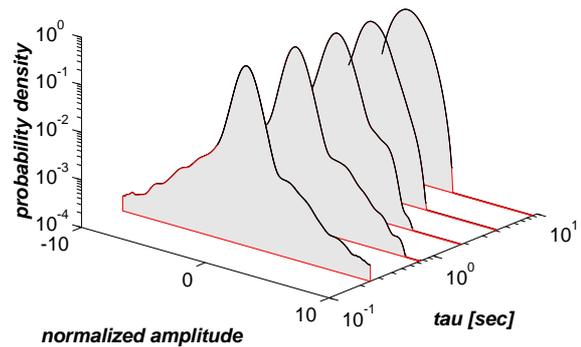

**Fig. 5.** The probability distribution function of the wavefield difference $\delta B(\tau) = B(t+\tau) - B(t)$ for the $B_x$ component of the magnetic field, in the upstream region. The distributions are centered and all have unit variance.

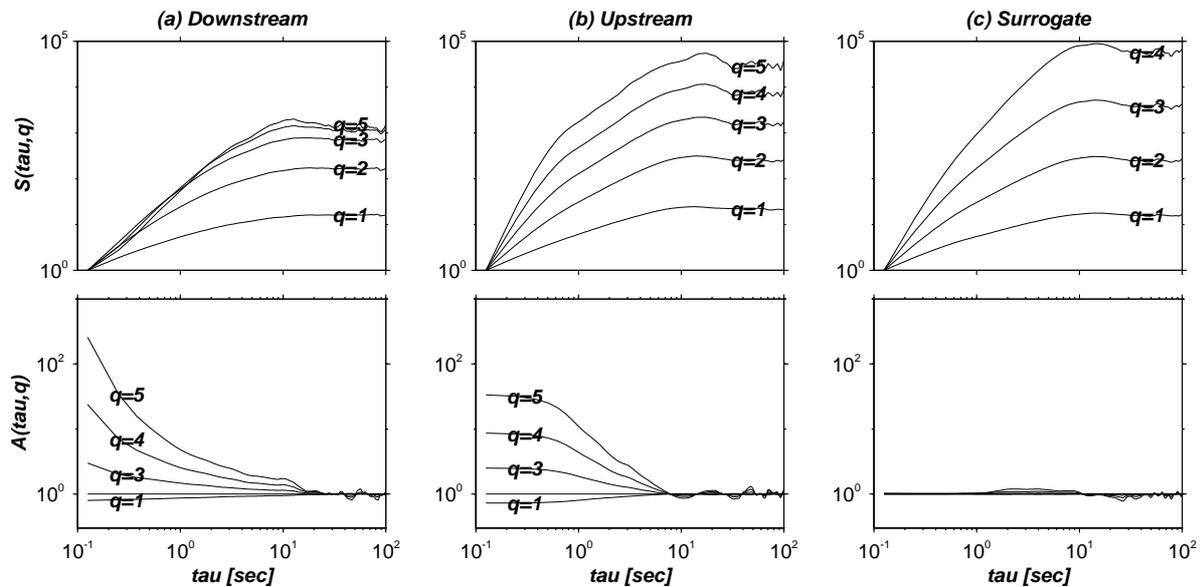

**Fig. 6.** The dimensional (upper row) and dimensionless (lower row) structure functions in the time domain for the $B_x$ component of the magnetic field. Three data sets are considered: (a) the downstream region, (b) the upstream region and (c) a surrogate of the upstream region that is Gaussian at all scales. Note that the time scales $\tau$ are expressed in seconds, with a sampling period of T=0.125 sec. For each order q, we have normalized the dimensionless structure functions to that of a Gaussian signal, so that $A_q(\tau) \equiv 1$ if there is no intermittency. Moreover, the dimensional structure functions are normalized with respect to $S_q(\tau=T)$.

## 4 Intermittency and fractal properties

A direct consequence of the observed intermittency is the lack of self-similarity, in the sense that a scaling of the type $S_q(\tau) \propto \tau^{\alpha(q)}$ cannot be applied except for small ranges of delays. This is in contrast to solar wind experiments, where much longer self-similar (i.e. inertial) ranges were found (Burlaga, 1992; Marsch and Liu, 1993; Carbone et al., 1996). We have nevertheless attempted an estimation of the scaling exponents in the range $1.2 < \tau < 5$ sec., which roughly corresponds to the frequency range in Fig. 3 in which both regions share the same the spectral index $\gamma \cong 2.4$. Clearly, there is some arbitrariness in the selection of the regression range. More robust criteria can be obtained by invoking the extended self-similarity hypothesis (Benzi et al., 1993), which states that scalings of the type $S_q(\tau) \propto S_p(\tau)^{\beta(q,p)}$ should hold for long ranges of delays $\tau$. Figure 7 indeed reveals that a $S_q(\tau) \propto S_3(\tau)^{\beta(q)}$ dependence holds relatively well from $\tau \cong 0.6$ sec. up to the largest delays under consideration.

The scaling exponents $\alpha(q)$ that are measured from the structure functions are shown in Fig. 8. In spite of their uncertainty one may safely conclude that $\alpha(q)$ departs from a linear function of q. A linear scaling is only recovered for surrogate data because the latter are self-similar by construction. This departure of $\alpha(q)$ from a linear law provides a priori a clear indication for intermittency, which may be quantified that way. To this end, we have computed the fractal dimension D, which reveals a more pronounced intermittent behaviour upstream ($D \cong 2.82 \pm 0.04$) than downstream ($D \cong 2.91 \pm 0.04$) the bow shock. The fact that D is between 2 and 3 suggests that the turbulence consists of a variable mixture of planar structures and three-dimensional eddies.

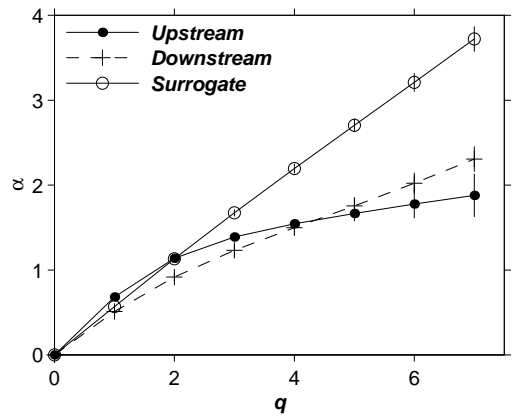

**Fig. 8.** The scaling exponents $\alpha(q)$ as measured in the range $1.2 < \tau < 5$ sec. from the up-, downstream and surrogate data of Fig. 6. The error bars reflect the uncertainty that comes from the inadequacy of the power law fit.

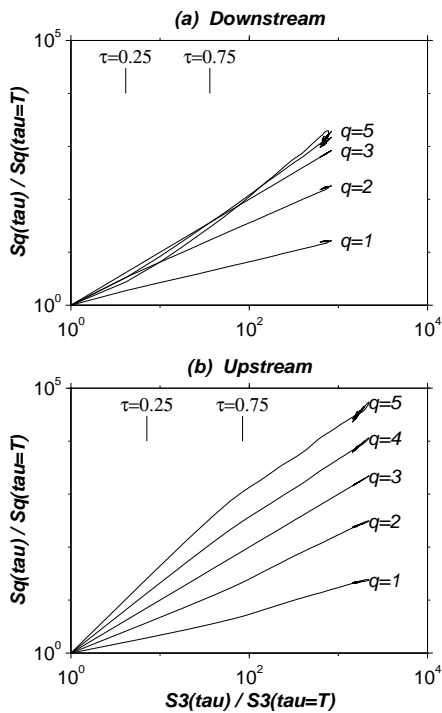

**Fig. 7.** The dependence of $S_q(\tau)$ on the third order structure function $S_3(\tau)$. The power law dependence over a large range of scales $\tau$ attests the existence of extended self-similarity. The latter is more evident in the downstream (a) than in the upstream (b) region. For visualization purposes, all the structure functions have been normalized to $S_q(\tau = T)$ where T=0.125 sec. is the sampling period.

The departure of scaling exponents from a linear law has received considerable attention in relation to underlying models of intermittency (Paladin and Vulpiani, 1987; Carbone, 1993; 1994; Grauer et al., 1994), all of which predict a more or less strong negative deviation of $\alpha(q)$ from a linear law. It remains questionable, however, whether such models can be identified unambiguously on the basis of our experimental data. Indeed, their discrimination requires high order moments whose computation becomes increasingly sensitive to data sampling effects. Some of these often overlooked sources of error are the lack of stationarity (Sethia and Reddy, 1995), the finite dynamic range of the instruments and the finite sample size (Tennekes and Wijngaard, 1972). We are particularly concerned here with the latter.

To assess the possible impact of finite sample size effects on our results, we apply a simple test that was proposed earlier by Tennekes and Wijngaard (1972). The idea is to consider for different orders q the distribution of the

magnetic field difference $b(\tau) = |B(t+\tau) - B(t)|$ and plot versus it the function

$$m_q(b) = b^q\, p(b), \qquad (10)$$

where $p(b)$ is the probability density of $b(\tau)$. The area which is spanned by this function is nothing but the structure function of interest since

$$S_q(\tau) = \int_0^\infty m_q(b)\, db. \qquad (11)$$

The integrands $m_q(\tau)$ are shown in Fig. 9 for the upstream region and with different values of q.

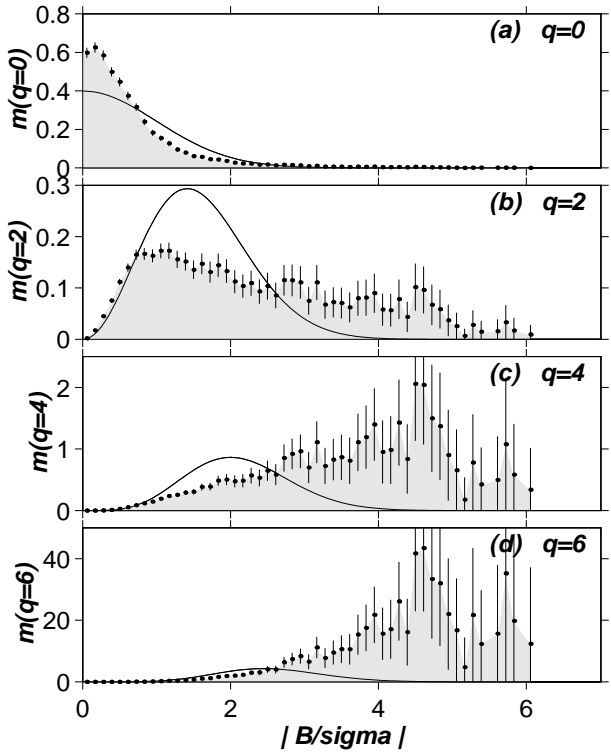

**Fig. 9.** Representation of the integrands $m_q(b)$ for the upstream data, with four different orders q and $\tau = 2$ sec. The gray areas correspond to the q'th order structure functions $S_q(\tau)$ and the full line to a reference Gaussian that has the same (unit) variance as the data. Note how the areas are truncated at large q because of the absence of large-amplitude events.

For low orders, the areas enclosed by these functions are relatively well defined since $m_q(\tau)$ decreases rapidly to zero. For q>2, however, the behaviour of the data points becomes increasingly erratic. This problem is not so severe insofar it merely increases the uncertainty of the estimates. A more pernicious effect occurs at high orders when the lack of large-amplitude events does not guarantee anymore reliable estimates of the area. A systematic underestimation of the structure function results, which introduces a negative bias in the scaling exponents. Unfortunately, this correction mimics the deviation one would expect from a multifractal process, thereby making any distinction between real multifractality and numerical artifacts very difficult.

The maximum order below which the results remain accurate within, say 10%, depends on the departure from Gaussianity. In our case, this limit can be approximately set at q=3 upstream and q=4 downstream, which means that fourth order moments are probably the highest ones we can reasonably expect to estimate from our data. Thus, all we can conclude so far is the increasingly intermittent behaviour of the wavefield at small scales, with a more pronounced effect in the upstream region.

## 5 Isotropy

Our conclusions regarding the interpretation of the higher order statistical properties of the wavefield may at this point look pessimistic. There are nevertheless some interesting issues. One of these is the isotropy of the non-Gaussian properties. Most theories on turbulence tacitly assume isotropy but this property turns out to be often, if not always violated in magnetofluids (Belcher and Davis, 1971). The main reason for this is the presence of a prevailing magnetic field that breaks the symmetry of the turbulence.

To investigate the isotropy of our data, we project the magnetic field along different directions and determine how evenly its statistical properties are distributed. The different orientations are parameterized by the elevation $-90° \leq \theta \leq 90°$ and the azimuth $-180° \leq \varphi \leq 180°$ in spherical coordinates. For each $(\varphi, \theta)$ pair, we consider $\hat{B}(\varphi, \theta)$ defined as

$$\hat{B} = \mathbf{B} \cdot \mathbf{p} \qquad (12)$$

with

$$\mathbf{p} = (\cos\varphi \cos\theta,\ \sin\varphi \cos\theta,\ \sin\theta)^T.$$

In these new coordinates, $B_x$ corresponds to $\theta = 0$ and $\varphi = 0$. Figure 10 summarizes some relevant statistical properties of the magnetic field using these spherical coordinates. Figure 10a displays the average magnetic field $<\hat{B}(\varphi, \theta)>$ and reveals a prevailing orientation with a maximum amplitude that is parallel to the solar wind upstream the shock and then deflected by about 45° after the shock crossing. The relative fluctuation amplitude of the wavefield is represented in Fig. 10b in terms of the standard deviation normalized to the total field strength

$$\sigma = \frac{\left\langle \left(\hat{B} - \langle\hat{B}\rangle\right)^2 \right\rangle^{1/2}}{\langle |\mathbf{B}| \rangle} \ . \quad (13)$$

Figure 10b reveals that the large-amplitude events (including nonlinear magnetic structures such as shocklets) are not regularly distributed but instead are concentrated in a plane. This attests their two-dimensional nature, a property that is well known for incompressible MHD turbulence (Matthaeus et al., 1990). Figure 10b thereby vividly illustrates the anisotropy of the wavefield. In the upstream region, the plane containing large-amplitude fluctuations is almost perpendicular to the prevailing magnetic field whereas it is oblique downstream. The former result is typical for Alfvén waves while the latter attests the lack of colinearity in the downstream region between the magnetic field and its gradient.

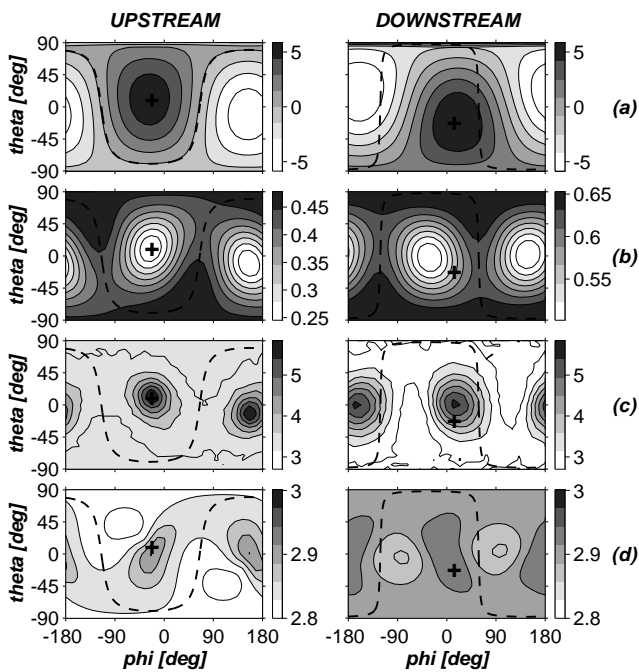

**Fig. 10.** Spatial distribution versus azimuth $\varphi$ and elevation $\theta$ of some statistical properties measured upstream (left column) and downstream (right column) the bow shock. From top to bottom are (a) the average magnetic field in nanoTeslas, (b) the relative fluctuation amplitude normalized to the total field amplitude, (c) the decorrelation time in sec. and (d) the fractal dimension that quantifies the departure from Gaussianity. Note that the vertical scales are pair-wise the same in plots (a), (c) and (d). For both regions we also display the prevailing orientation of the magnetic field (marked by a cross) and the plane that contains the largest fluctuation amplitudes (dashed line).

The dynamic properties of the wavefield are partly summarized in Fig. 10c in terms of the decorrelation time or integral scale $\tau_B$. This quantity expresses the loss of "memory" of the wavefield and is defined here as the time it takes for the autocorrelation function to drop from 1 to 0.5. The fact that $\tau_B$ varies significantly suggests that by changing the orientation we also pick up different types of waves. Maxima in the decorrelation time clearly coincide with extrema in the mean magnetic field and are likely to be related to compressible magnetosonic-like waves that oscillate along the magnetic field. Small decorrelation times in turn coincide with Alfvén-like waves that oscillate transverse to the prevailing magnetic field. Thus, not only are the static but also the dynamic properties of the wavefield irregularly distributed.

Finally, we consider the local statistical properties of the wavefield and to this end compute the fractal dimension D from Eq. (5). The choice of this parameter is arguable but it has the merit to be dimensionless and thereby eases the comparison. The most important point in Fig. 10d is not so much the interpretation of the fractal dimension as its irregular distribution. The strongest deviations from Gaussianity appear in the upstream region where they coincide with regions that exhibit large standard deviations, see Fig. 10b. Downstream the shock, the fractal dimension is more isotropic and closer to its upper limit 3. Such a dependence of the higher order moments on the orientation was already noticed by Marsch and Liu (1993); we now find that it is correlated with the macroscopic properties of the magnetic field. Indeed, the least intermittent behaviour (i.e. the values of D closest to 3) occurs parallel to the prevailing magnetic field, and thereby corresponds to small amplitude magnetosonic-like waves. Conversely, the strongest deviations from Gaussianity coincide with the location of the nonlinear magnetic structures. These results support our conceptual picture in which nonlinear waves such as shocklets are more likely to affect higher order statistical moments than linear magnetosonic-like waves which propagate along to the magnetic field.

In view of these results, it is evident that the statistical properties of the turbulent wavefield are anisotropic. The main reason for this is the coexistence of several types of waves that proceed from different physical mechanisms and therefore differ in their statistical properties. The interpretation of such a mixture in terms of intermittent processes is not obvious and quantitative comparisons with theoretical models are certainly not appropriate. Note, however, that the different types of waves are relatively well separated in the upstream region, which is likely to be the reason for its higher anisotropy. The bow shock acts on the turbulence as a mixing process that makes the wavefield more stochastic and thereby smears out the intermittent features. Such a heuristic picture would explain the more regular distribution observed in the downstream region.

## 6 Spectral statistics of the wavefield

Since the statistical properties of the wavefield are likely to receive different contributions, an important issue is to find additional means for disentangling its constituents. One possible solution is a spectral representation. The structure function has the property to offer excellent time resolution but it cannot resolve different frequencies, which are known to be more adequate for describing wave phenomena. We are therefore motivated to consider a different representation of the structure function that offers improved frequency resolution. First note that the structure function can be rewritten as

$$S_q(\tau) = \left\langle \left| B_W(t,\tau) \right|^q \right\rangle, \qquad (14)$$

where $B_W(t,\tau)$ stands for a wavelet transform of B(t) at a scale $\tau$

$$B_W(t,\tau) = \int_{t_1}^{t_2} h\left(\frac{t'-t}{\tau}\right) B(t')\, dt' . \qquad (15)$$

This expression normally contains an additional scaling in $\tau^{-1/2}$ to preserve energy conservation. The function

$$h(t) = \delta(t-1) - \delta(t) \qquad (16)$$

may be interpreted as a special kind of backward difference wavelet; it has zero mean but does not possess the admissibility conditions that are required for wavelet transforms (Farge, 1992). Its time resolution is excellent, but its high pass characteristics cause a spectral leakage that precludes a resolution of waves with closely spaced frequencies. We replace it by a real Morlet wavelet (also known as a Gaussian wavelet) that offers a better compromise between time and frequency resolution

$$h(t) = Re\left(\frac{1}{\pi^{1/4}} e^{-2\pi j t} e^{-t^2/2}\right). \qquad (17)$$

Inserting this into Eqs. (14) and (15) gives us a different type of structure function $\tilde{S}_q(\tau)$ whose time scale $\tau$ is now the inverse of an instantaneous frequency f. The spectral structure function may be interpreted as the q'th order moment of the original data after applying a bandpass filter centered on $f = 1/\tau$. $\tilde{S}_4(\tau)$, for example, is directly related to the kurtosis of the filtered wavefield. A major advantage is its ability to probe higher order moments in terms of frequencies and not just time delays. This property was already advocated by Mahrt (1991) for atmospheric turbulence and has been successfully used in the analysis of multifractal processes (Muzy et al., 1994).

A brief comparison between the two types of structure functions raises some interesting points that are summarized in Fig. 11. Both representations reveal the same deviation from Gaussianity at small scales. This agreement is not surprising since it merely confirms that for a randomly fluctuating wavefield, the choice of the wavelet has no strong bearing on the results as long as a proper common time base is used. This is not true anymore when the turbulence contains coherent oscillations. The reason for this is that Morlet wavelets are much more efficient in capturing coherent waves than are usual structure functions, which smear them out. An example of such a coherent wave is given in Fig. 12, where it stands out against the randomly fluctuating background by its deterministic behaviour.

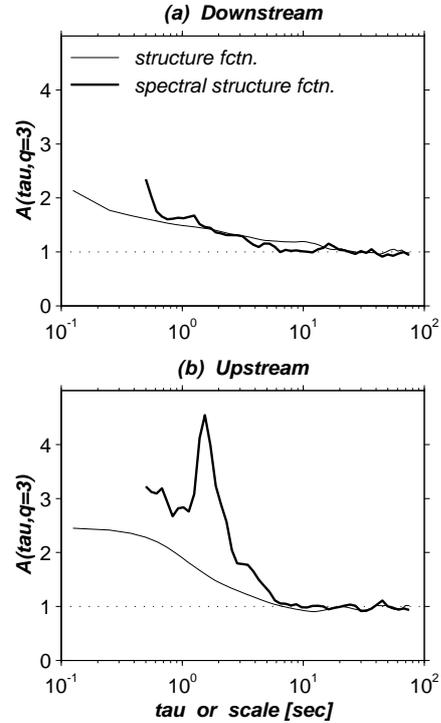

**Fig. 11.** The third order (q=3) dimensionless structure functions of the $B_x$ component measured downstream (a) and upstream (b), using standard and spectral structure functions. The latter clearly reveal the existence of coherent waves upstream the bow shock at $\tau \cong 1.6$ sec., corresponding to a frequency of 0.6 Hz. The normalization is the same as in Fig. 6.

Several coherent wave packets can be identified in our data. Their frequencies all fall between 0.55 and 0.7 Hz, which exactly corresponds to the peak of the spectral structure function in Fig. 11. These almost monochromatic waves have been identified as whistler waves that grow out of the large-amplitude shocklets through a competition between nonlinearity and dispersion (Dudok de Wit et al., 1995). These whistlers occur on both sides of the shock but they are more apparent upstream where Landau damping is less efficient. It is interesting to note that their relative energy content is small enough to keep the Fourier spectrum devoid of spectral lines (see Fig. 3). Yet, their intermittent occurrence is easily detected by the spectral structure function. We conclude that a comparison between temporal and spectral structure functions can be highly informative by revealing the presence of coherent oscillations that otherwise would go undetected.

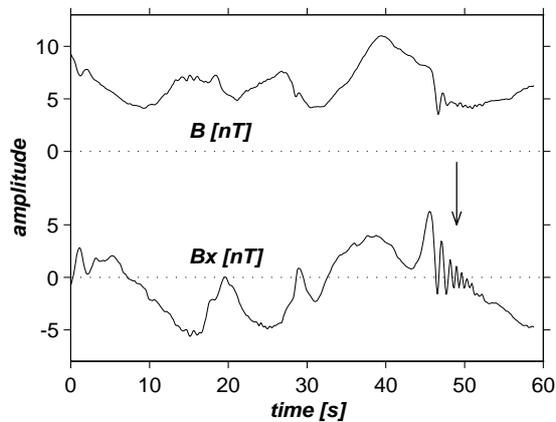

**Fig. 12.** Excerpt of the wavefield showing the time evolution of the total magnetic field (top) and its $B_x$ component (bottom) in the upstream region. A typical whistler wavetrain is indicated by an arrow; this wave packet results from the decay of the large-amplitude structure that precedes it at t=40 sec.

The existence of whistler waves in our case study reveals another caveat in the interpretation of structure functions. Clearly, we are in the presence of two different types of intermittency which we cannot distinguish. The wavetrain in Fig. 12 is intermittent because it occurs in sudden bursts that occur erratically. This type of intermittency is reminiscent of that found in dynamical systems (e.g. Eckmann, 1981). The notion of intermittency that tacitly underlies models on small-scale turbulence in contrast supposes spatial or spatio-temporal spottiness. These two different types of intermittency are difficult to separate unless they can be identified by other means such as by spatio-temporal measurements. This problem is also related to the distinction between local and global self-similarity, as discussed by Vassilicos (1993). It reveals the danger of misinterpreting the intermittent behaviour of a fluctuating wavefield when the spatio-temporal properties cannot be clearly established. It thereby highlights the importance of adding spatial resolution in order to properly understand plasma turbulence.

Given the presence of whistler waves upstream the bow shock, we now have to reconsider our results in the light of the possible impact of coherent waves. To this end, we have recomputed the structure functions of Fig. 6 while skipping all sequences that contain whistler waves. The resulting subset has a power spectral density which hardly differs from that of the original data; the higher order statistical properties, however, are substantially different. This difference is particularly evident in the distribution of the fractal dimension D which is now more akin to that of a Gaussian process, see Fig. 13. Indeed, the measured value of D is now much closer to its upper limit D=3 one would expect for Gaussian scale-invariant turbulence. We conclude that the intermittency observed in the upstream region is more due to the occasional occurrence of whistler wavetrains than to some intrinsic intermittency of the inertial range, which itself is close to Gaussian. The same conclusion probably holds downstream the shock, but we have not attempted any correction because the whistlers are not as easy to isolate.

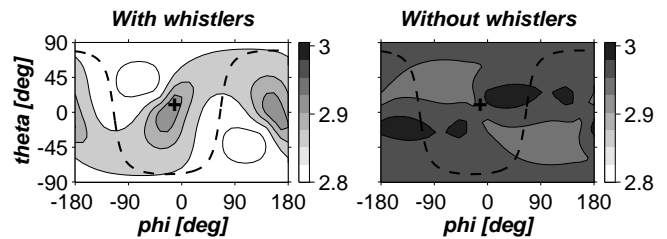

**Fig. 13.** Spatial distribution versus azimuth $\varphi$ and elevation $\theta$ of the fractal dimension estimated in the upstream region using raw data (left) and after removing sequences that contain whistlers (right). Removing a few whistlers causes the turbulent wavefield to become statistically much closer to a Gaussian isotropic process, for which D=3 everywhere.

## 7 Conclusions

The initial objective of this study was a comparison between higher order statistical properties of plasma turbulence as it undergoes a sudden compression at the Earth's collisionless bow shock. The structure functions of the fluctuating magnetic field are qualitatively comparable to what has previously been observed in solar wind MHD turbulence, with a departure from Gaussianity at small time-scales that attests the presence of intermittency. Both sides of the shock reveal similar properties which a priori suggests that the small-scale properties of the turbulence are little affected by the sudden transition. Such an interpretation, however, suffers from several caveats.

We first note that the statistical properties strongly depend on the orientation of the wavefield, thereby attesting its small-scale anisotropy. The strongest deviations from a Gaussian distribution appear upstream the shock, transverse to the prevailing magnetic field, and coincide with the orientation of the large-amplitude nonlinear magnetic structures termed shocklets. Conversely, the least intermittent behaviour occurs along the magnetic field for magnetosonic-like waves.

Another significant effect is the nonlinear dispersion of the wavefield. This dispersion invalidates the Taylor hypothesis and precludes the usual one-to-one connection between the observed temporal dynamics and predicted spatial structure. Dispersive effects are not exceptional in plasma turbulence, yet they're often neglected by lack of experimental support. Such an assumption may seriously compromise a quantitative assessment of the scaling exponents in space plasmas.

The observed statistical properties are also found to be deeply affected by the occasional occurrence of coherent whistler wavetrains. These discrete wave packets hardly contribute to the total power spectral density, yet they substantially increase the departure from Gaussianity. By subtracting them from the data, a more isotropic and much less intermittent wavefield is recovered, whose properties are similar up- and downstream the bow shock. The intermittent signature we observe is more due to a temporal intermittency that is reminiscent of dynamical chaos than to the type of spatio-temporal inhomogeneity which is usually understood. A straightforward interpretation in terms of fractal properties is therefore not possible.

Finally, we note that the finite duration of our records prohibits a reliable estimation of structure functions beyond the fourth order. This pernicious and yet often overlooked effect leaves substantial freedom in the interpretation of the scaling exponents.

After revealing several pitfalls in the interpretation of higher order moments, is there still any reason to believe that they're are as appropriate for plasma turbulence as they are for hydrodynamic flows ? Indeed, most of the properties that make them useful for neutral fluid turbulence are apparently much less relevant to plasma turbulence, in which their interpretation is less obvious. There remain, however, several interesting issues. Obviously, a separation of the wavefield into the physically more relevant Elsässer variables (e.g. Biskamp, 1993) should be used when possible, since this eases the interpretation. We have also highlighted the importance of considering the spatial structure of the wavefield in addition to the temporal dynamics. Another issue is the spectral representation of the structure function, that has been introduced here and is relevant for investigating wave-like patterns in turbulence. Finally, we note that most of the effort in modelling cascade processes has gone so far into statistical models. Alternative approaches, such as closer connections with the theory of dynamical systems (Ugalde and Lima, 1996) may provide new insight.

*Acknowledgments.* The AMPTE-UKS magnetometer data were kindly provided by M. Dunlop (Imperial College, London). We would also like to thank an anonymous referee for helpful suggestions. Financial support by the Swiss National Science Foundation and the Institut National des Sciences de l'Univers during an early part of this work is greatly acknowledged.